\documentclass[Afour,sageh,times, doublespace]{sagej}

\usepackage{moreverb,url}

\usepackage[colorlinks,bookmarksopen,bookmarksnumbered,citecolor=red,urlcolor=red]{hyperref}

\newcommand\BibTeX{{\rmfamily B\kern-.05em \textsc{i\kern-.025em b}\kern-.08em
T\kern-.1667em\lower.7ex\hbox{E}\kern-.125emX}}

\def\volumeyear{2020}

\begin{document}

\runninghead{M Byrne}

\title{Simple post-translational circadian clock models from selective sequestration}

\author{Mark Byrne}

\affiliation{Physics Department, Spring Hill College,
Mobile, AL  36608  USA}


\email{mbyrne@shc.edu}

\begin{abstract}
It is possible that there are post-translational circadian oscillators that continue functioning in the absence of negative feedback transcriptional repression in many cell types from diverse organisms. Apart from the KaiABC system from cyanobacteria, the molecular components and interactions required to create in-vitro ("test-tube") circadian oscillations in different cell types are currently unknown. Inspired by the KaiABC system, I provide "proof-of-principle" mathematical models that a protein with 2 (or more) modification sites which selectively sequesters an effector/cofactor molecule can function as a circadian time-keeper. The 2-site mechanism can be implemented using two relatively simple coupled non-linear ODEs in terms of site occupancy; the models do not require overly special fine-tuning of parameters for generating stable limit cycle oscillations.
\end{abstract}


\maketitle

\section{Introduction}
There are circadian ($\approx 24$ hr) clocks in most organisms \cite{dunlap1999molecular}. 
These internal biological clocks are experimentally characterized by sustained oscillations in one or more measured outputs under (approximately) constant conditions (e.g., constant light or darkness).  Circadian oscillators can also be entrained via external stimuli, such that the internal oscillator's period and relative phase can be synchronized by the entraining stimulus \cite{roenneberg2003art}.  In contrast to general biochemical oscillators, the oscillatory period of a circadian clock remains almost unchanged for different constant temperatures in the relevant physiological range (temperature compensation) \cite{hong2007proposal}.  The simplest known example of a circadian clock that functions outside cells (in a "test-tube") is the 3-protein KaiABC system \cite{tomita2005no, nakajima2005reconstitution}.  This type of clock is termed a post-translational oscillator (PTO) since its operation does not require dynamical feedback from a genetic circuit. In contrast, the fundamental oscillatory design of the circadian clock in most organisms in vivo is believed to consist of a transcription-translation negative feedback loop (TTFL) which is influenced by post-translational modifications (see, for example \cite{young2001time,lee2001posttranslational,lowrey2000genetics}).  However, negative feedback TTFLs do not generally imply sustained oscillations \cite{kurosawa2002comparative, qin2010coupling}.  In this context, mathematical models were originally required as proof-of-principle that the TTFL mechanism was quantitatively viable, in the sense that simple TTFL mathematical models could, in principle, reproduce experimentally realistic circadian oscillations in the relevant measured outputs \cite{goldbeter1995model}.  

Both simplified (2D) and more complex mathematical models of eukaryotic TTFL circadian clocks have been investigated \cite{tyson1999simple,forger2003detailed}.  However, there is evidence of the existence of circadian PTOs in a variety of species absent genetic feedback loops \cite{o2011circadianA, o2011circadianB, edgar2012peroxiredoxins} .  These findings suggest that it may be useful to investigate post-translational designs which can satisfy the defining conditions for a circadian clock \cite{jolley2012design}. The elucidation of basic molecular designs which can create circadian clocks can presumably serve as a guide to experimental searches for possibly other protein-based "in vitro" PTOs and for investigating the apparent degree of simplicity (or difficulty) required for intracellular post-translational molecular methods of timing.  

In particular finding simplified two dimensional mathematical models of circadian clocks is potentially useful since a variety of mathematical tools (phase plane analysis, Hopf bifurcation theory) can be used to analyze and visualize the system's dynamics. This study demonstrates the existence of two simple designs whose mathematical models are perhaps the simplest 2D circadian clock models for post-translational molecular oscillators. Both designs require a protein with at least two regulatory sites, selective sequestration of an effector molecule by one of the protein states, and a separation into fast-slow kinetics for regulation of site occupancy. These general conditions are sufficient to generate stable limit cycle oscillations in the population occupancy of the sites for reasonable rates and parameter values.  Furthermore this class of “selective sequestration” models permits both entrainment and temperature compensation, by assuming one of the sites (the "fast" site) is sensitive to external perturbation while the "slow" regulatory dynamics on the other site is essentially "buffered" from external perturbations.

\section{Results}

\subsection{Post-translational oscillatory mechanisms from two protein sites and selective sequestration}

For simplicity assume a "core" clock protein ($X$) with two modification sites ($a = 1$ and $b = 2$) in which the rate of modification of the residues can be altered by another "effector" molecule ($Y$).  For the sake of generality we suppose the molecule that is transferred to each site of the protein remains unspecified (e.g., a phosphoryl group, ATP, or an oxygen atom) - for the KaiC protein there are two phosphorylation sites per KaiC monomer and the rate of (auto) phosphorylation of the sites is affected by another protein (KaiA) in a hyperbolic manner \cite{iwasaki2002kaia, rust2007ordered}.  We propose two general simple designs for an autonomous clock using sequestration of $Y$ based on the occupancy of the two sites (Fig 1).

\begin{figure}
\centering
\includegraphics[width=100mm]{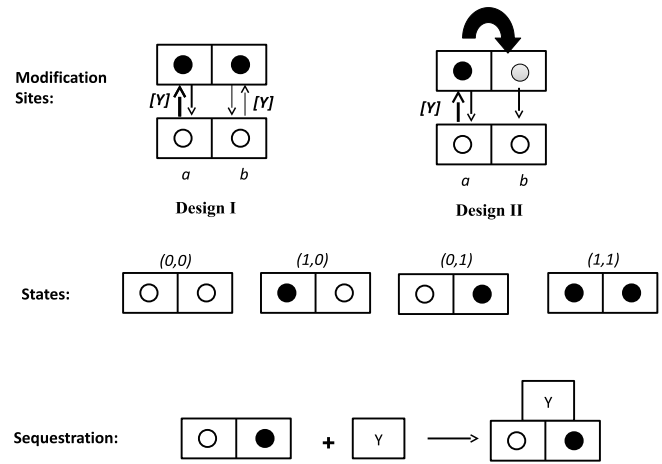}
\caption{
Two possible simple designs for a protein-based clock. In design I, two modification sites on the protein (not necessarily adjacent) are both regulated by [Y]. Filled circles indicate occupancy of a protein site, by addition of a small molecule to that residue (for example). In design II the occupancy of one site (regulated by [Y]) is transferred to a second site. Only one of the four protein states sequesters the effector molecule, Y.)
}
\end{figure}

In one oscillatory scheme each site is independently modified and the kinetics proceed at different rates on the two sites; in a second class of models the occupancy of one site is transferred to the other site via some unspecified mechanism (e.g., an intra-protein transfer).  Assume the population kinetics of the two sites follows 1st order kinetics so that the fractional occupancy of each site ${(0 < x_{i} < 1}$, e.g., degree of “phosphorylation”) in the population is described as follows:

\begin{equation}{
	dx_{i}/dt =  k_{i} (1-x_i) - k_{-i} x_{i}	}	\qquad (i = 1,2) 
\end{equation}

where $i$ labels each modification site and a constant time-independent “decay” (e.g., “dephosphorylation”) term is allowed in the kinetics.  In general the modification rate(s) on the site(s) are assumed to vary as smooth, monatonic functions, $f$, of the effector concentration, which we parameterize as $k_{i} = k_{i,max} f([Y])$ where $0 < f([Y]) < 1$; hyperbolic and linear regulatory functions of the rates are examined below.

\begin{figure}
\centering
	\includegraphics[width=120mm]{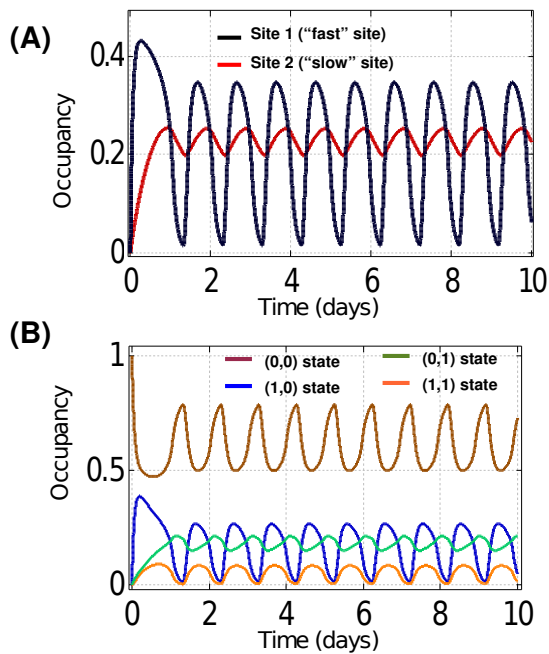}
\caption{Design I with independent (hyperbolic) modification of the sites by [Y].  (A) Example population occupancy kinetics for the two protein sites (rates and parameters are in the main text). (B) Dynamics for the corresponding 4 protein states.}

\end{figure}

\begin{figure*}
\centering
	\includegraphics[width=80mm]{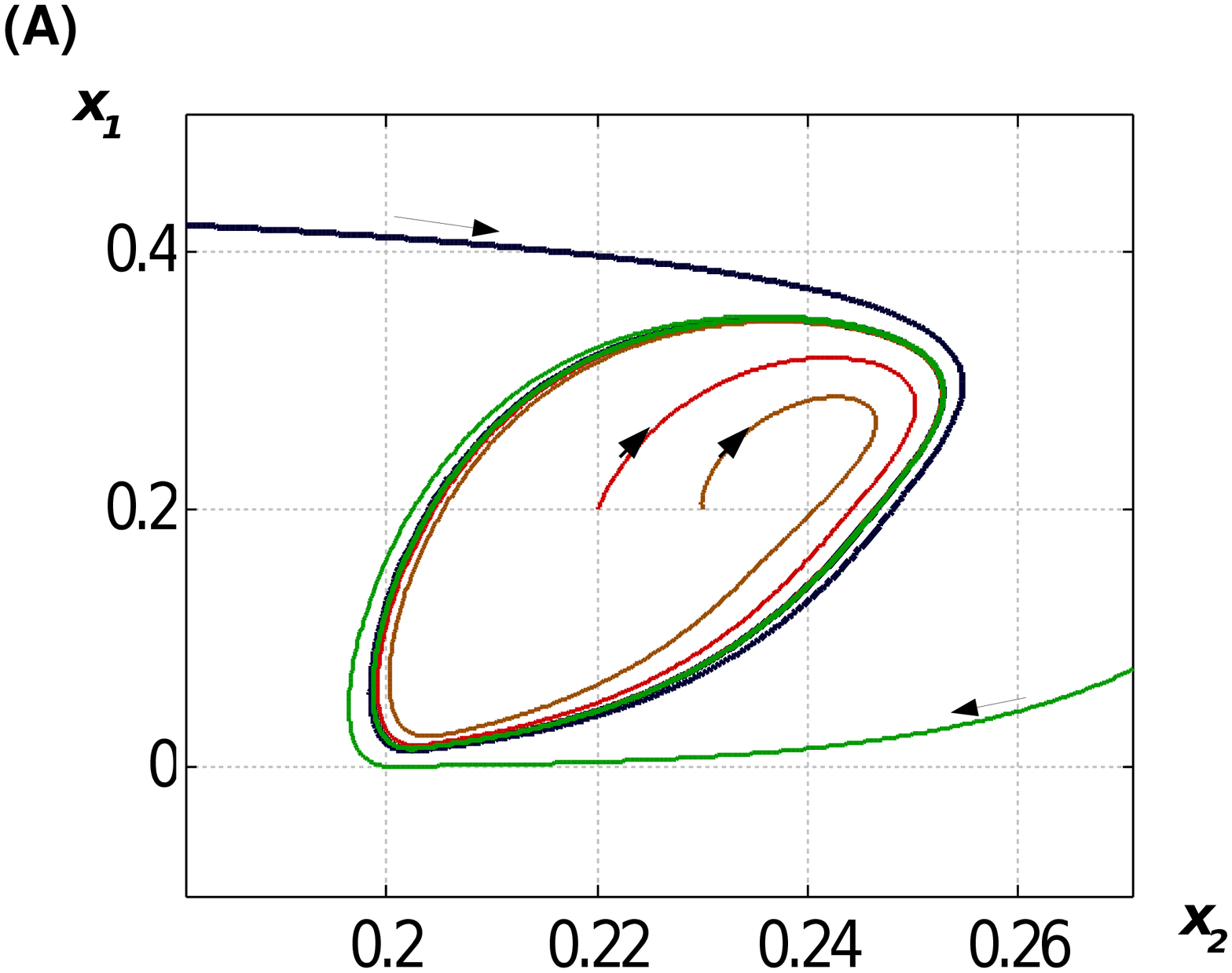}
	\includegraphics[width=80mm]{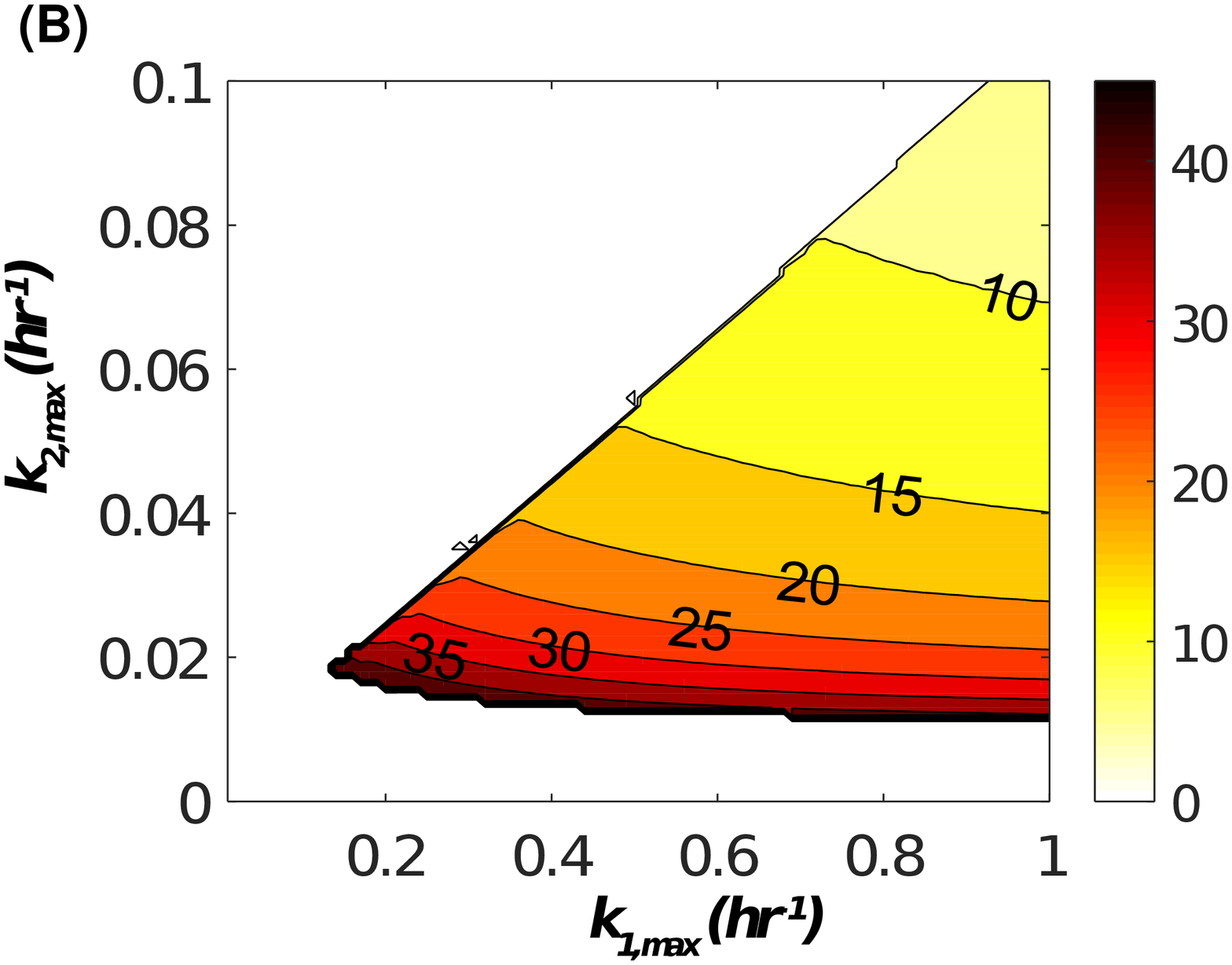}
	\includegraphics[width=80mm]{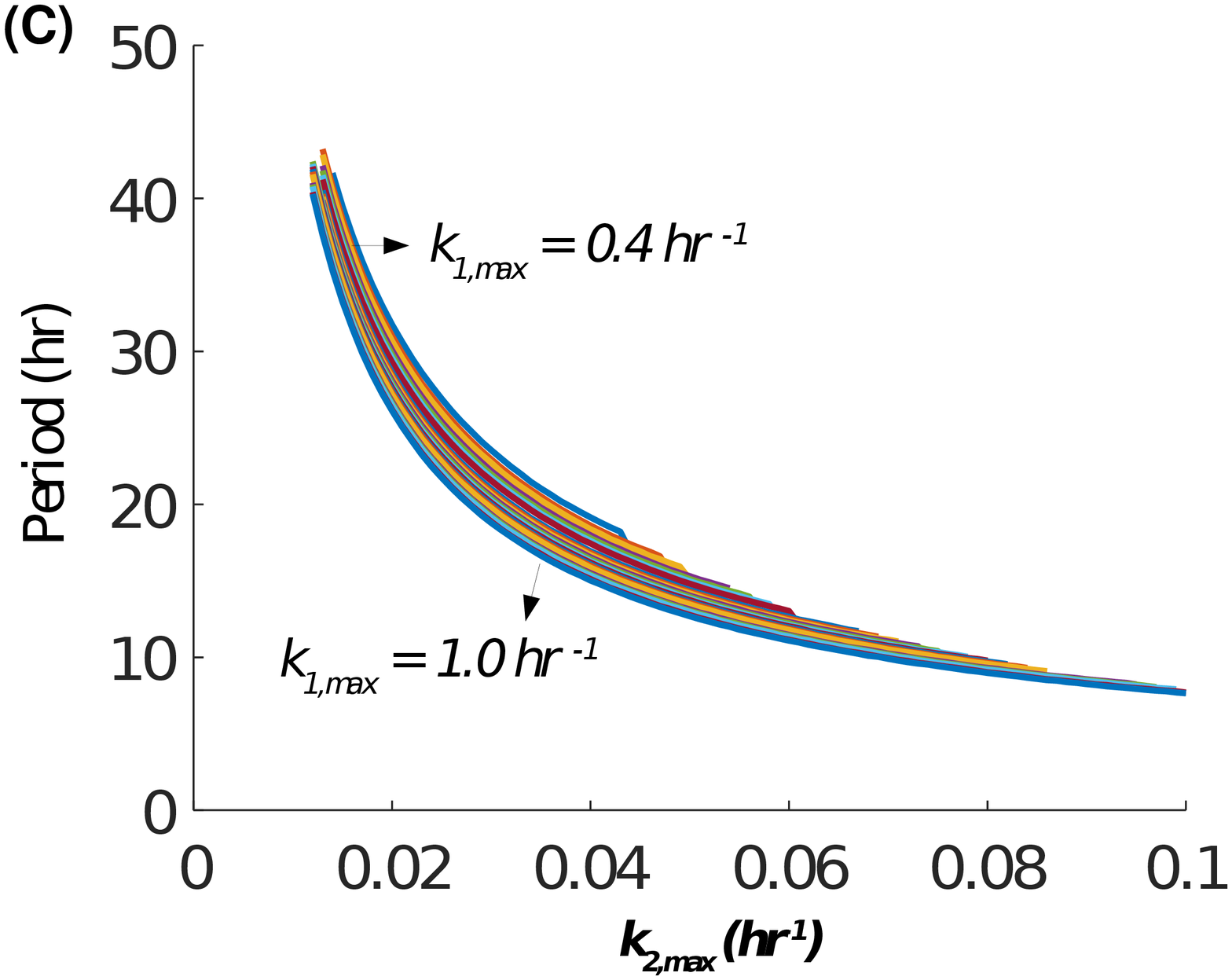}
	\includegraphics[width=75mm]{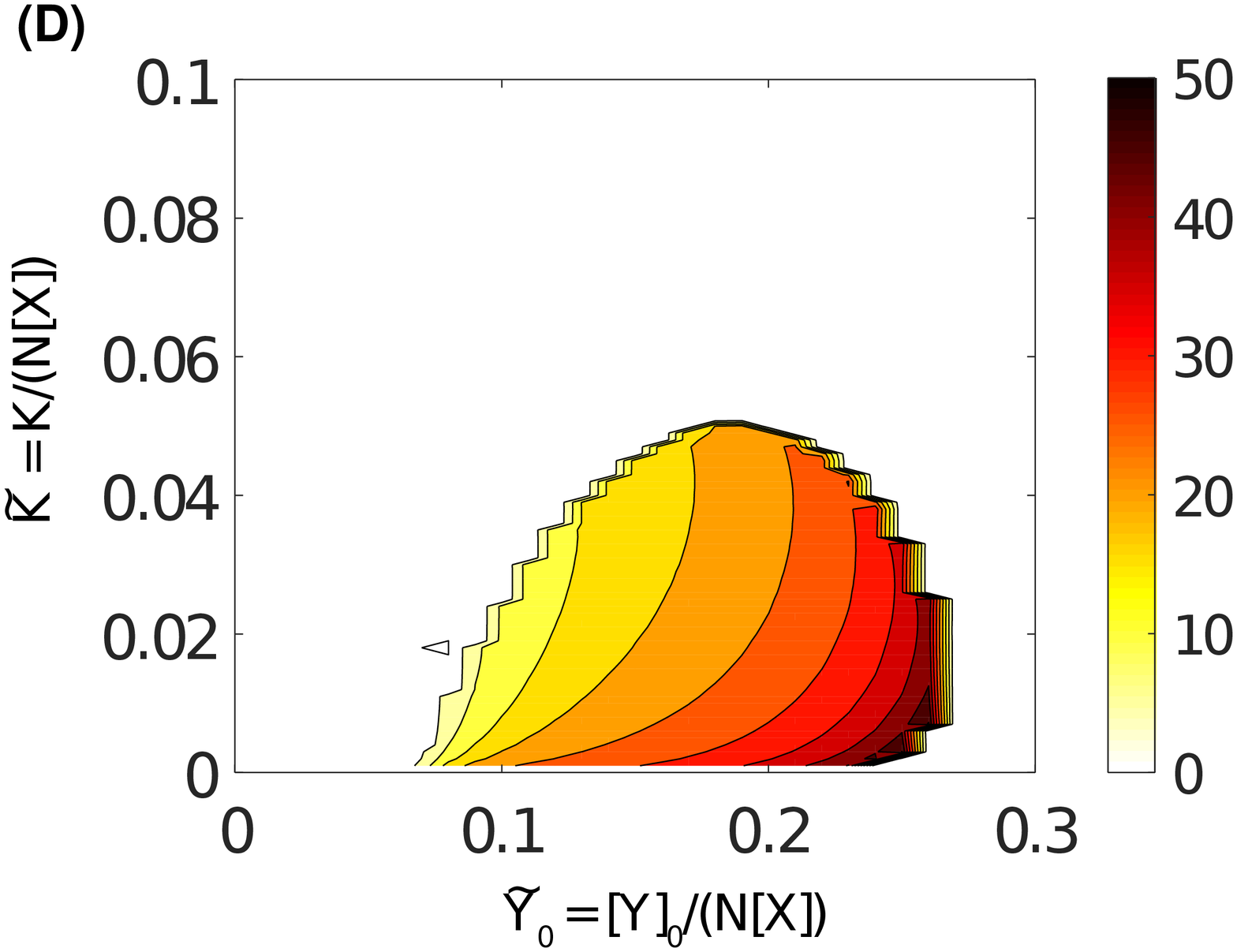}
\caption{Model I: (A) Stable sample limit cycle trajectories for different initial conditions using the rates and parameters from fig 2. (B) Parameter region allowing oscillations and the oscillatory period in hours assuming $\tilde{Y_0}  = 0.20, \tilde{K}  = 0.04$ (C) An elaboration of Panel (B), indicating sensitive dependence of period (power-law) to the "slow" rate constant for site 2 modification (D) Parameter region allowing oscillations and the oscillatory period in hours for a given dimensionless Michaelis constant and scaled initial concentration of effector, assuming $k_{1,max} = 0.4 hr^{-1} = k_{-1}$,  $k_{+2,max} = 0.03 hr^{-1} = k_{-2}$.}
\end{figure*}

An individual protein ($X$) with two modification sites can be in one of four states: $(a,b) = (0,0),(1,0),(0,1),(1,1)$ with zero indicating an un-occupied site and one occupied. Population statistical arguments imply the average fraction of proteins in each of the four states is 
	
\begin{equation}
\begin{split}
	& x_{(0,0)} = (1- x_1) (1-x_2) \\
	& x_{(1,0)} = x_1 (1-x_2) \\
	& x_{(0,1)} = x_2 (1-x_1) \\
	& x_{(1,1)} = x_1 x_2
\end{split}
\end{equation}

Now suppose the modification rates ($k_{i}$) of both sites follow Michealis-Menten regulatory kinetics, depending hyperbolically on the concentration of the effector molecule $(Y)$:  

\begin{equation}{
	k_{i} = k_{i,max} \frac{[Y]}{[Y] + K}.
	}	
\end{equation}
Letting site $a$ represent the 'fast' site ($ k_{1, max} >> k_{2,max}$) breaks the symmetry in the model and sequestration of $Y$ by the $(0,1)$ protein states can generate sustained oscillations depending on the rate constants and model parameters.  A simple "rigid" model of sequestration is to instantaneously alter the concentration of $Y$ according to the concentration of $X$ proteins in one of the four protein states \cite{rust2007ordered}. If an average of $N$ molecules of $Y$ are sequestered per $X$ protein in the state $x_{(0,1)}$, the non-sequestered concentration of $[Y]$ a time $t$ is given by:

\begin{equation}{
	[Y] =  max\{ [Y]_{0} - N[X] x_{(0,1)} ,0\}}
	\label{eq:four}
\end{equation}

where $[Y]_0$ indicates the initial concentration of molecule $[Y]$.  For this mechanism, the dependence of the system dynamics in terms of the model parameters and rates is best seen by rescaling equations (2) and (4) by the fixed protein concentration $[X]$ and the parameter $N$:  
 
\begin{equation}
	k_{i} = k_{i,max} \frac{\tilde{Y}} {\tilde{Y} +\tilde{K}}
\label{eq:five}	
\end{equation} 
and
\begin{equation}
	\tilde{Y} = max\{\tilde{Y_0} -  x_2 (1-x_1),0 \}
	\label{eq:six}
\end{equation}  

where  $\tilde{Y} \equiv \frac{[Y]} {N [X]}$ is the dimensionless scaled fraction of the effector, $Y$;  $\tilde{Y_0} \equiv \frac{[Y_0]} {N [X]}$ is the equivalent dimensionless scaled fraction of initial effector $[Y_0]$, and $\tilde{K} \equiv \frac{K}{N [X]} $  is a dimensionless Michaelis constant. The system dynamics is encoded by the two fixed parameters (${\tilde{Y} }_0$ and  $\tilde{K}$) and three (ratios) of rate constants (since one rate constant can be absorbed into a dimensionless time parameter by re-scaling both differential equations in Eqn.(1) by, e.g., $1/ k_{1,max}$).  For example, sustained $\approx 24$ hr oscillations are reproduced using the parameters $\tilde{Y_0}  = 0.20, \tilde{K}  = 0.04$ and the rates $k_{1,max} = 0.4 hr^{-1} = k_{-1}$,  $k_{2,max} = 0.03 hr^{-1} = k_{-2}$  (Fig 2). Numerical integration suggests a limit cycle upon varying the initial conditions (Fig 3A). Sustained oscillations are possible as a result of an autocatalytic negative feedback loop (from sequestration), previously suggested in several mathematical models of the KaiABC clock \cite{clodong2007}, \cite{van2007}, \cite{rust2007ordered}. In this 2D model the qualitative description for sustained oscillations is that the slow kinetics on one site allows the site with rapid kinetics to reach near maximal occupancy before sequestration takes effect. Once sequestration starts to occur from the slow formation of $(0,1)$ states, loss of occupancy on the fast site drives the transition $(1,1) \rightarrow (0,1)$. This causes further sequestration of $Y$ and increases the rate of further loss of occupancy on the fast site creating additional $(0,1)$ states (auto-catalysis).  Slow loss of occupancy from the 2nd site assures that the 1st site becomes largely unoccupied until de-sequestration of $Y$ occurs simultaneous with the transition $(0,1) \rightarrow (0,0)$. Numerical investigations of the solution space of this model (Eqs. 1,4 and 5) indicate that the existence of oscillations is quite sensitive to the parameters $\tilde{Y_0}$ and $\tilde K$ (Fig 3D). Qualitatively as the relative concentration of initial effector is lowered (for constant $\tilde{K}$) the oscillatory period decreases because sequestration of $Y$ takes less time; for fixed initial effector, increasing the Michaelis constant delays the onset of sequestration resulting in a longer oscillatory period. Numerical solutions indicate that sustained oscillations require a "fast-slow" separation of timescales for the maximum modification rates on the two sites (Fig 3B). Varying both $k_{1,max}$ and $k_{2,max}$ over the range $[0,1]$  $hr^{-1}$ in steps of $0.01$ $hr^{-1}$ (and setting $k_{-1} = k_{1,max}$,  $k_{-2} = k_{2,max}$) suggests sustained oscillations occur when $k_{2,max} < 0.1 k_{1,max}$ (Fig 3B). As expected the oscillatory period is only weakly dependent on $k_{1,max}$ but strongly dependent (power-law dependent with a power $\approx$ -0.75) on $k_{2,max}$ (Fig 3C), which is important in interpreting both temperature compensation and entrainment in these clock models.

Interestingly there is an even simpler class of sequestration models that can generate stable limit cycles (and might be employed biochemically). Consider the "transfer" design in which the molecule occupying site 1 is transferred to site 2 at rate $k_T$ so that the model is now described by a class of ODEs:

\begin{equation}
\begin{split}
	&  dx_{1}/dt =  k([Y]) (1-x_1) - k_{T} x_{1} - k_{-1} x_{1} 	\\	
	&  dx_{2}/dt =  k_{T} x_1 - k_{-2} x_{2}	
\end{split}	 
\label{eq:seven}
\end{equation}

where terms for both transfer and loss of occupancy for site 1 have been included. 
In these designs the rate $k$ can depend linearly on the substrate ($k = k_{+1} \tilde{Y}$) and  generate limit cycle oscillations (hyperbolic variation can also generate stable limit cycles). Non-dimensionalizing ($\tau \equiv {k_{+1} t}$) and neglecting loss of occupancy without transfer (for the moment), $k_{-1} = 0$,  gives the following simple system ($\alpha  \equiv \frac{k_{T}}{k_{+1}}, \beta \equiv \frac{k_{-2}}{k_{+1}}$), where :

\begin{equation}
\begin{split}
	& f(x_1,x_2) \equiv dx_{1}/d\tau =  \tilde{Y}(1-x_1) - \alpha x_{1}	\\	
	& g(x_1,x_2) \equiv dx_{2}/d\tau =  \alpha x_1 - \beta x_{2}
\end{split}	 		 
\label{eq:five}
\end{equation}

with $\tilde{Y}$ given by Eqn.(5). For example there are stable limit cycle oscillations for  $\alpha = 0.25, \beta = 0.1$, $\tilde{Y_0} = 0.5$; setting $k_{+1} = 0.7hr^{-1}$ generates a circadian timescale for the oscillations (Fig 4, 5A).  Allowing loss of occupancy from site 1 lowers the oscillation amplitude and can result in damped oscillations; examples for $k_{-1} = 0.01 hr^{-1}$ and $k_{-1} = 0.05 hr^{-1}$ are shown in Fig 5B.  

\begin{figure}
	\includegraphics[width=100mm]{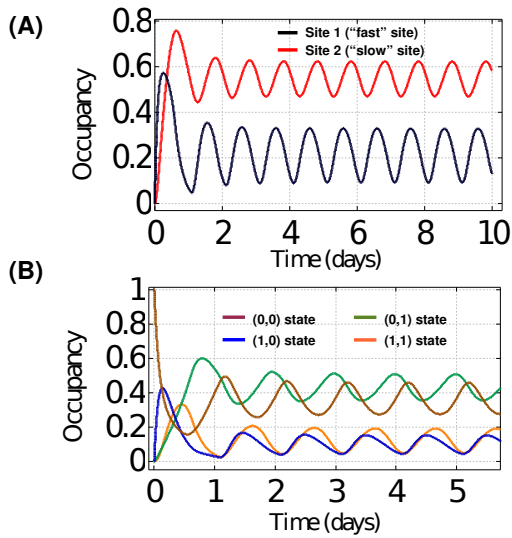}
\caption{Design II with transfer of occupancy from site 1 to site 2.  (A) Example population occupancy dynamics for the two protein sites (rates and parameters are in the main text). (B) dynamics for the corresponding four protein states.}
\end{figure}

\begin{figure}
\centering
	\includegraphics[width=100mm]{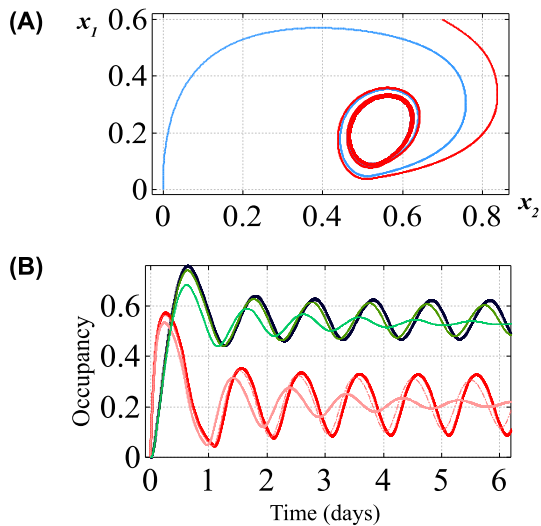}
\caption{Design II: (A) Stable sample limit cycle trajectories for different initial conditions using the rates and parameters from fig 4. (B) Allowing both loss of occupancy from site 1 and transfer to site 2; $k_{-1} = 0.05 hr^{-1}$ (thin trace, damped oscillations) and $k_{-1} = 0.01 hr^{-1}$ with slightly reduced amplitudes compared to Fig 4A traces (thicker red and black)}
\end{figure}

\begin{figure}
\centering
	\includegraphics[width=80mm]{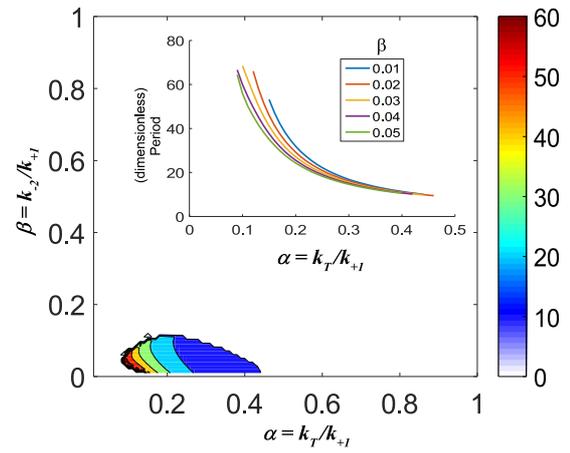}
\caption{Approximate region of stable limit cycle oscillations for design II from numerical integration; colormap shows dimensionless period as the ratio of rates are varied (fixed $\tilde{Y}_{0} = 0.5$) and the inset shows the approximate power-law dependence of the dimensionless period on $\alpha$.}
\end{figure}

The simple dynamical system can be studied analytically  - see Appendix for a linear perturbative analysis about the steady-states. Direct integration of the ODEs confirms the regions of instability and oscillatory period estimates from a linear stability analysis of the dynamical system (see Appendix Fig 9). As an example, consider $\tilde{Y_0} = 0.5$. Linear (in)stability constraints suggest oscillations generally require $\beta < 0.1, \alpha < 0.5$ and $\beta < \alpha$ (Fig 6). The dimensionless oscillatory period ($\tilde{P}$) is fit well ($R^2 > 0.99$) by different power-law functions for a given $\beta$ : $\tilde{P} \approx A \alpha ^{B}$ with $(A,B) = (2.74,-1.53)$ for $\beta = 0.01$; $(A,B) = (3.37,-1.23)$ for $ \beta= 0.05$; and $(A,B) =  (2.89, -1.27)$ for $\beta= 0.1$.  The physical period is $P = \frac{ \tilde{P}}{k_{+1}}$ which, in this regime of parameters for oscillations gives an approximate physical oscillatory period: 

\begin{equation}
P \approx A  {k_{T}}^{B} {k_{+1}}^{- B -1} 
\end{equation}

For example, for $\beta= 0.1$: 
\begin{equation}
P \approx 2.89~ k_{T}^{-1.27} k_{+1}^{0.27}   
\end{equation}

Thus the oscillatory period in this class of models is primarily set by the "transfer" rate ($k_{T}$). 

\subsection{Entrainment and Temperature Compensation}
	
Circadian clocks are both sensitive to external perturbations (can be entrained) yet retain near period invariance under varied environmental conditions, such as temperature fluctuations. In these simple models we can suggest a possible mechanism; the oscillator period is "mildly" sensitive to external perturbations so that the oscillation period is approximately constant and set by the "slow" internal dynamics, while the "fast" dynamics (e.g., $k_{+1}$) incorporates external perturbations. In this model oscillator, all perturbations (metabolic, light or dark pulses, temperature, etc.) interact only through the modification rates, assuming the perturbation does not alter the relative protein abundances ($\frac{[Y_0]}{[X]}$). In these models the change in period ($\delta P$) in terms of any perturbation in rates ($\delta k_j$) is, to 1st order, 

\begin{equation}
\delta P = \sum_{j} \frac{\partial{P} }{\partial{k_j}} \delta{k_j} 
\end{equation}

This perturbation should be approximately zero for "compensation" to be effective (temperature or other perturbations).  There are two generic possibilities; one is that the effective rates in the model are "trivially" insensitive to the perturbation ($\delta{k_j} \approx 0$) due to structural properties of the protein(s). However if the system were completely "trivially" compensated then entrainment would not be possible (except by protein and/or effector abundance variation). The other is that the system is "tuned" to some degree so that one perturbation which tends to increase the period is countered by another that decreases the period \cite{Ruoff_1992}). More generally, in the parameter space of $P(k_j)$ approximately flat regions correspond to approximate period-invariant sub-spaces of the parameter space. The approximate power-law dependence in these models (Eqn 9) displays this compensatory mechanism with the constraint that the dynamical system's period remains invariant when setting the sum of partials to zero in the power-law approximation: 

\begin{equation}
\frac{\delta k_T}{k_T} = (1+ 1/B)  \frac{\delta k_{+1}}{k_{+1}}
\end{equation}

\begin{figure}
\centering

	\includegraphics[width=120mm]{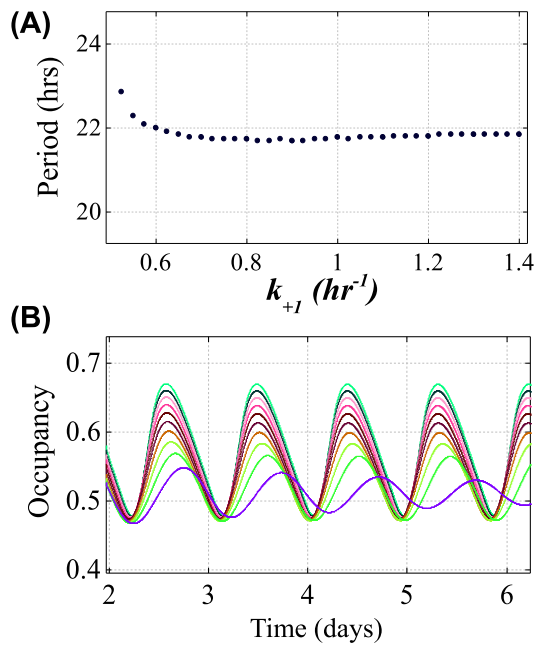}
	
\caption{(A). Approximate period invariance of the oscillations as the fast rate ($k_{+1}$) is varied while the transfer rate ($k_T$) is slightly adjusted according to Eqn (11) in the text with $B \approx -1.4$. (B). Sample site 2 occupancy oscillations corresponding to the variation of $k_{+1}$ shown in Panel A (larger oscillation amplitudes correspond to larger $k_{+1}$). }
\end{figure}

For example, consider initial unperturbed rates of $k_{+1} = 0.7 hr^{-1}$ and $ k_T = 0.2 hr^{-1}$ for $\beta= 0.04$. Numerically the period is $P \approx 21.77 hrs$. Simulations suggest a doubling of the site 1 rate (typical for a 10 $^{\circ}$C increase), $k_{+1} = 1.4 hr^{-1}$ is almost compensated ($P_{new} \approx 21.86 hrs$) by about a $30$ percent increase in the transfer rate (corresponding to $B \approx -1.4 $, $ k_T = 0.26 hr^{-1}$), Fig 7A. It is clear from direct integration of the model that period compensation is possible as predicted by Eqn 11 (Fig 7A,B); as the fast-rate is further lowered below $k_{+1} \approx 0.5 hr^{-1}$, the oscillations start damping with increasing period. If we assume the rate doubling on the 1st site corresponds to a 10 $^{\circ}$C increase, these parameters give a Q10 (for the period) of about $1.004$. Since these are tuned parameters, a desired Q10 can be selected by a judicious choice of rate compensation, corresponding to a choice of activation energy threshold(s) for regulation of site occupancy using the Arrhenius-Boltzmann temperature-dependence of the rates ($k_i \propto A_i \exp(-{E_i}/(kT)$), as previously described for general biochemical kinetics (\cite{Ruoff_1992}). 


Entrainment in the model is examined using both continuous and discrete (pulse) perturbations. For small amplitude perturbations, "stable" entrainment is possible within an approximate range of 19 to 26hrs for these parameters. The following external driving function was assumed: $\frac{\delta k_{+1}}{k_{+1}} = A sin( 2 \pi t / \tau)$ with amplitude, $A =0.25$ chosen. The transfer rate was modified to retain near period invariance in the absence of the external continuous sinusoidal perturbation ($B \approx -1.4$ in eqn 11). For example, the period shifts from $21.77 hrs$ (unperturbed) to $24.02 hrs$ for an external $\tau = 24$ hr rhythm and from $21.77$ hrs (unperturbed) to $19.04$ hrs for an external $\tau = 19$ hr rhythm (Fig 8A).  Entrainment is also examined using 2-hr pulse reductions in the rates. Phase response curves (PRCs) are constructed using 2-hr pulses by transiently reducing the the on-rate starting on day 4.5 of the unperturbed oscillator (Fig 8 B,C). The sample PRCs in Fig 8 show reductions of $k_{+1} = 1.0 hr^{-1}$ to $\{0.0, 0.5, 0.8\}$ $hr^{-1}$ starting on day 4.5 of the unperturbed oscillator over one oscillations cycle and computing the phase shift relative to the unperturbed control on day 10 of the oscillation. The transfer rate was also adjusted as previously described according to eqn 11 using $B = -1.4$. The long "dead phase" of the PRC in these models is because sequestration already abrogates the fast rate ($k_{+1} \approx 0$) so that further reductions in the rate by an externally applied down-pulse in this rate has little effect during the time interval of significant sequestration. 

\begin{figure*}
\centering

	\includegraphics[width=120mm]{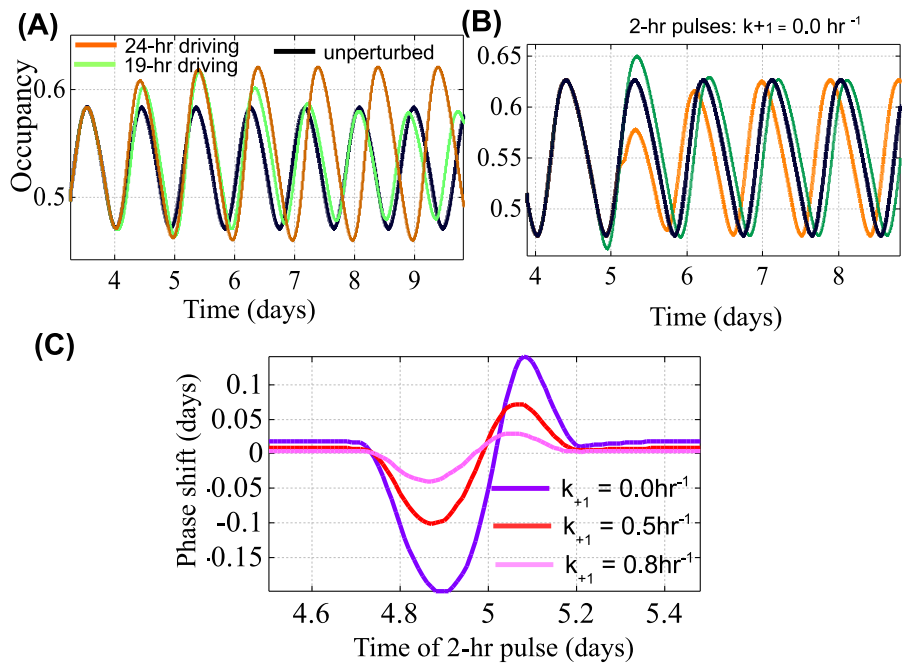}
	
\caption{(A). Entrainment to externally driven oscillations of the rate $k_{+1}$ starting on day 4; black = control, orange = 24hr, green = 19hr. (B). Sample traces including pulse reductions in the fast rate (setting $k_{+1} = 0 $ for 2-hr intervals starting on day 4.5). (C). Sample phase response curves (PRCs) for varying amplitude pulse reductions (measured on day 10) with $k_T$ varied according to eqn 11.}
\end{figure*}

\section{Discussion}
The model designs and simulations in this paper suggest that the molecular interactions required for a post-translational circadian clock could be surprisingly simple. In these designs, selective sequestration of an effector (regulator) of protein site occupancy and a separation of time-scales in the dynamics of regulation of the protein sites appear to be sufficient for generating sustained oscillations. The model can be easily generalized to a protein with $N$ regulatory sites of which a subset of the $2^N$ states selectively sequesters an effector (or multiple effectors) for finer clock regulation, coupling to other proteins, improved "buffering" of the slow dynamics, etc. Already, with just 2 protein sites and 4 states it is possible to both entrain the oscillations to external cues and have the limit cycle oscillations compensate for fluctuations in the fast dynamical variable (e.g., from temperature or metabolic perturbations). A limitation of the model is that much complexity is encoded in the effective rate constants, including energy regulation and ATP-ADP dynamics which were not included in these models to keep the parameter space as small as possible. In particular for circadian clocks, the effective rates are much slower than typical rates from enzyme or binding kinetics ($s^{-1}$ or less), whereas TTFL models incorporate a natural several $hr$ timescale that is part of the transcription-translation process. In the KaiABC clock, the slow ATPase dynamics of KaiC are implicated in the slow characteristic oscillatory timescale (\cite{Terauchi_2007}) and a similar slow ATP hydrolysis would likely be involved in setting the effective rates in these models. Another limitation in these models, also used to reduce the potential state-space of dynamical variables, is that sequestration was not explicitly modeled using mass action (for example); the linear sequestration model by one of the protein states (eqn 5) is an effective and direct method to simulate selective sequestration (\cite{rust2007ordered}) but introduces a hard cutoff in these models reflected in the fast rate abruptly shifting to zero.    
  
A prediction of the simplest 2-site/4-state version of the model is that evolutionary mechanisms should have selected rates such that the existence of oscillatory behavior is robust under typical fluctuations of the effective rates; the "transfer" model predicts the most likely values of $\alpha$ in the range $0.1$ to $0.4$ and small $\beta$ ($< 0.1$). Design I with hyperbolic regulation suggests "fast" effective modification rates of $0.5$ to $1.0 hr^{-1}$ and a "slow" modification rate less than $1/10$ of these values. Presumably, similar to the KaiABC clock, slow conformational dynamics of the protein(s) structure is implicated in the slow regulatory (and period-determining) dynamics. A further prediction of these models is the rather general approximate power-law dependence of the oscillation period on the slow rate, which might be tested using various clock mutants with "cloistered" regulatory site(s). Perhaps these and similar models will assist and encourage the experimental search for in-vitro protein-based oscillators beyond the remarkable KaiABC cyanobacterial clock. 

\section{Methods}
Numerical integration of the ODEs was implemented in Fortran (GNU Fortran G77, Free Software Foundation) using 4th-order Runge-Kutta. The sequestration constraint (eqn 5) was implemented using a threshold of $0.001$ for $\tilde{Y}$. In-house fortran code was written for scanning parameter space, peak finding, estimating oscillatory periods (typical uncertainty $< 0.05 hr)$, and applying perturbations (fig 8). For figures 7 and 8, the initial conditions, $x_1(t = 0) = 0.5 =x_2 (t =0)$ were used. Other parameters are listed in the main text. Colormap figures (Fig 3B,D, Figs 6,9) were generated using Matlab (The MathWorks, Inc., Natick, Massachusetts, United States).

\subsection{Appendix: Linear Perturbative Analysis of Simple Transferase Model}

\begin{figure}
\centering

	\includegraphics[width=100mm]{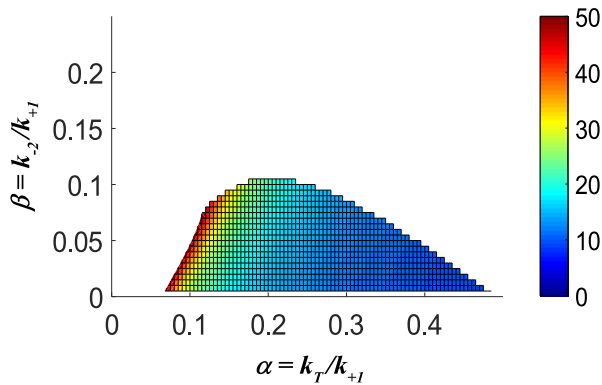}
	
\caption{Predicted oscillatory parameter space and dimensionless oscillatory period (colormap) for design II using analytical solutions for the fixed point(s) of the dynamical system (eqn 7) and applying the conditions for (linear) instability of the steady-state(s) from the matrix, $M$.}
\end{figure}

In this section a linear perturbative analysis about the steady-state solutions of the design II models (Eqn 7) is used to suggest constraints on the parameters that permit (or do not permit) limit cycle oscillations. The fixed points ($x^{*}_{i} \ni dx^{*}_{i}/d\tau = 0$) are solvable analytically (from a cubic) and are functions of two effective parameters only, $\beta$ and $\tilde{Y_0} \frac{\beta}{\alpha}$. A linear perturbation about the steady-states yields a stability matrix, evaluated at the steady state(s):

\[ M = \left( \begin{array}{ccc}
2 x^{*}_{2} (1-x^{*}_{1})- (\tilde{Y_0} +\alpha) & -1 + x^{*}_{1}(2-x^{*}_{1})  \\
\alpha & -\beta 
\end{array} \right)\]   

Potential oscillatory solutions with instability of the steady state (unstable spiral) requires $Tr M > 0$ and $Det M > \frac{1}{4} (Tr M)^2$. Since $x^{*}_{2} = \frac{\alpha}{\beta} x^{*}_{1}$ and $max\{(1-x^{*}_{1})x^{*}_{1}\} =1/4$ this implies

\begin{equation}
0 < \beta (\frac{\tilde{Y_0}}{\alpha} + \frac{\beta}{\alpha} + 1) < 1/2
\end{equation}

Since each term is positive this implies $\beta < 1/2$. Further constraints are given using Bendixson's theorem; the sum of partial derivatives is
\begin{equation}
\frac{\partial f}{\partial x_1}+ \frac{\partial g}{\partial x_2} = -( \alpha  + \beta + \tilde{Y_0} ) + 2x_2 (1-x_1). 
\end{equation}

As the latter quantity is confined on the interval $[0,2]$ assuming the domain $0 < x_i <1$ this sum of partials will be strictly negative (and thus not permit limit cycles) unless  
\begin{equation}
0 < \alpha + \tilde{Y_0} < 2,
\end{equation}

Since each parameter is positive, $\alpha < 2, \tilde{Y_0} < 2$. Direct numerical evaluation of the steady states and applying the conditions for instability as functions of $\alpha$ and $\beta$ give tighter ranges of instability consistent with these analytical constraints (Fig 9). The imaginary component($\omega$) of the eigenvalues of $M$ (for each unstable steady-state from solving the cubic and applying the instability criteria) gives a "local" perturbative approximation to the dimensionless period ($\tilde{P} =\frac{2\pi}{\omega}$) as functions of $\alpha$ and $\beta$ (see Fig 9 colormap). The estimate of the linear perturbative analysis is in good agreement with the numerical estimate of the oscillatory parameter space and corresponding oscillation periods from direct numerical integration of the ODEs.

\bibliographystyle{SageH}
\bibliography{mybib}

\begin{acks}
I appreciate comments and discussion on early work on this project from Carl Johnson, Tetsuya Mori, Ximing Qin and Yao Xu (Vanderbilt U.). I also express my appreciation to Spring Hill College (and especially Dr Lesli Bordas) for enabling a "sabbatical" leave during the Spring 2020 semester.
\end{acks}

%
%
%

\end{document}